\documentclass[letter]{aa}

\usepackage[latin1]{inputenc}
\usepackage[T1]{fontenc}
\usepackage{moreverb}
\usepackage{fancyhdr}

\usepackage{graphicx}
\usepackage{amsmath}

\usepackage{natbib}
\bibpunct{(}{)}{;}{a}{}{,}

\usepackage[english]{babel}

\usepackage{txfonts}

\begin{document}

\title{Mobility of D atoms on porous amorphous water ice surfaces under interstellar conditions}

\author{E.~Matar
\and E.~Congiu
\and F.~Dulieu
\and A.~Momeni\thanks{Present address: LCAM, Universite Paris-Sud, Orsay, France}
\and J.~L.~Lemaire}

\institute {Universit\'e de Cergy-Pontoise \& Observatoire de Paris,
LERMA, UMR 8112 du CNRS, 95000 Cergy-Pontoise, France\\
\email{Elie.Matar@u-cergy.fr}}

\offprints{E. Matar}

\date{Received ; accepted}

\abstract{} {The mobility of H atoms on the surface of
interstellar dust grains at low temperature is still a matter of
debate. In dense clouds, the hydrogenation of adsorbed species
(i.e., CO), as well as the subsequent deuteration of the accreted
molecules depend on the mobility of H atoms on water ice. 
Astrochemical models widely assume that H
atoms are mobile on the surface of dust grains even if controversy
still exists. We present here direct experimental
evidence of the mobility of H atoms on porous water ice
surfaces at 10~K.}
{In a UHV chamber, O$_2$ is deposited on a porous
amorphous water ice substrate. Then D~atoms are deposited onto the
surface held at 10~K. Temperature-Programmed Desorption~(TPD) is
used and desorptions of O$_2$ and D$_2$ are simultaneously monitored.}
{We find that the amount of O$_2$ that desorb during the TPD
diminishes if we increase the deposition time of D atoms. O$_2$
is thus destroyed by D atoms even though these molecules
have previously diffused inside the pores of thick water ice. Our
results can be easily interpreted if D is mobile at 10~K on the
water ice surface. A simple rate equation model fits our experimental data and best fit curves were obtained for a D atoms diffusion barrier of 22$\pm$2 meV. Therefore hydrogenation can take place efficiently
on interstellar dust grains. These experimental results are in line with most calculations
and validate the hypothesis used in several models.}
{}


\titlerunning{Mobility of D atoms on p-ASW at 10 K}

\maketitle

%----------------------  Introduction  -----------------------

\section{Introduction}\label{intro}

Among the numerous molecules detected in different
astrophysical environments, a large fraction is composed of
hydrogenated species \citep[and references therein]{chang2007}. 
It has been established that dust grains play
a major role in the hydrogenation of compounds \citep{herbst1973, herbstFD}. 
In the dense interstellar medium where most of the hydrogenated 
species have been detected (especially during the star
formation stage \citep{tegmark1997, cazauxtielens, cazauxspaans}), 
grains are covered in a molecular
mantle mainly composed of water ice. In all the theoretical chemical models that
describe the hydrogenation of compounds on grains \citep{tielens1982, cuppen2007}, mobility of
hydrogen is hypothesised, though some 
calculations on  amorphous water surfaces \citep{smoluchowski-1981} 
and modelling of experimental data \citep{perets2005} rule out such a mobility. 
On the other hand, other calculations \citep{buch1991, masuda1998}
and interpretations of experimental data \citep{hornekaer2003} validate
the hypothesis that H is mobile at 10~K, a temperature supposed to
be close to that of dust grains in interstellar dense clouds where
hydrogenation takes place. We have already demonstrated that D atoms
are mobile at 10 K (in our experiments time scale) on non porous 
Amorphous Solid Water (np-ASW) ices \citep{Amiaud2007}, 
but the porosity and the roughness of the water ice films grown at low temperatures
(10~K) may considerably lower the D atom mobility.

Yet the debate upon the mobility of hydrogen atoms is not closed,
and is widely nourished with experimental interpretation controversy
\citep{vidaliFD, dulieuFD}. If one considers only the latest estimated
value of 51 meV \citep{perets2005} for the diffusion energy barrier of H
on porous amorphous solid water (p-ASW) ice, the hopping time between two adjacent
adsorption sites is about 10~million years at 10~K. In such a
context, no hydrogenation chemistry can occur on icy mantles of dust
grains on a reasonable time scale. Hence, observational evidence of
hydrogenated and deuterated species is a good argument for an
experimental investigation of the mobility of H-atoms.

The aim of this paper is to provide new experimental facts that
directly address---without the prism of a sophisticated model---the
question of hydrogen mobility on p-ASW ice at 10~K. By
using O$_2$ as a tracer of D mobility we present a set of
experiments that are straightforward to interpret if D is mobile on
the surface of porous amorphous water ice, as it is expected from
almost all calculations and included in several astrochemical
models. In Section~\ref{sectexp} we briefly describe the
experimental set-up and procedures. In Section~\ref{sectres} we present
our experimental results and explain them assuming that D
is mobile on p-ASW ice at 10~K. In Section ~\ref{model}, we describe a simple rate equation model that we used to fit our experimental results. In Section~\ref{sectdisc} we discuss other interpretations before concluding.

%----------------------  Experimental setup  -----------------------

\section{Experimental Section}\label{sectexp}

%%%%%%%%%%%%%%%%%%%%%%%%%%%%%%%%%%%%%%%%%%%%%%%%%%%%%%%%%%%%%%%%%%

\subsection{Experimental Set-up}\label{sectapparatus}

The FORmation of MOLecules in the InterStellar Medium (FORMOLISM)
experimental set-up has been developed with the purpose of studying
the reaction and interaction of atoms and molecules on surfaces
simulating the dust grains under interstellar conditions (relevance
of substrate, low density, and very low temperature $\sim10$ K).
FORMOLISM is composed of an ultra-high vacuum chamber with a
base pressure of $\sim$10$^{-10}$~mbar, of a rotatable quadrupole
mass spectrometer (QMS) and of an oxygen-free high-conductivity
copper sample holder. The sample holder is attached to the cold 
finger of a closed-cycle He cryostat and can be cooled down to 8~K. 
The temperature is measured with a calibrated silicon diode clamped 
on the sample holder and controlled by a Lakeshore 334 controller 
to $\pm$0.2~K with an accuracy of $\pm$1~K in the 8--400~K range.
Reactants are introduced into the vacuum chamber via two separated triply
differentially pumped beam lines aimed at the surface. The first
one is used to deposit O$_2$ while the second beam line
is used to introduce the D atoms. This second beam line consists, in
its first stage, of an air-cooled quartz tube surrounded by a
microwave cavity for dissociating D$_2$. To cool the D
atoms exiting the dissociation region, the source is terminated with an
aluminum nozzle connected to a closed cycle He cryostat that can be cooled down
to $\sim25$~K. For the experiments presented here the D beam has
a temperature of 50~K and a flux of $\sim$1 10$^{13}$cm$^{-2}$. The dissociation rate is constant at
$\sim$50\%.

%%%%%%%%%%%%%%%%%%%%%%%%%%%%%%%%%%%%%%%%%%%%%%%%%%%%%%%%%%%%%%%%%%%%%%%

\subsection{Growth of the ASW films}\label{sectASW}

For this work we grow composite ice films of p-ASW on np-ASW. The
sub-layer of np-ASW has a thickness of
$\sim$50~ML \mbox{(1 ML = 10$^{15}$ cm$^{-2}$)} and is grown at
120~K with a micro-channel array doser. The only purpose of the
compact ice sub-layer is to isolate the ensuing p-ASW films from 
the Cu surface \citep{engquist95}. On top of the np-ASW substrate, the
p-ASW films are grown at 10~K. In the present
study we use two different thickness of p-ASW, namely 20~ML
and 250~ML. The 20~ML p-ASW film is grown from a background
flux formed by controlling the H$_2$O partial pressure in the vacuum
chamber. Since ASW films deposited from ambient vapour take a long time to
form and since a long time is needed to reach the base pressure again, the
250~ML p-ASW films are deposited with the micro-channel
array doser. A previous calibration has shown that under these 
conditions we grow 0.33~ML of ASW per second. Although p-ASW films 
grown by this method exhibit a lower degree of porosity, the larger 
thickness ensures a more complex pore structure than that of thin 
p-ASW ices grown by background dosing.
After the growth of the composite ice films, they are annealed to
70~K to avoid any subsequent morphological changes. Although the
porous structure is not completely destroyed by the annealing
process, this reduces irreversibly the porosity of the p-ASW films
\citep{kimmel-2001-2, kimmel-2001-1}. This is verified by looking at the shift in
temperature of the D$_2$ desorption peak as described in Hornekaer et al. \citeyear{hornekaer2005}.

\subsection{Measurement Procedures}\label{sectproc}

Our experimental procedures are the following:
\begin{itemize}
\item
After the preparation of a stable and pure O$_2$ beam, the
p-ASW surface held at 10~K (or 25~K in another set of experiments) is exposed to 0.5~ML of O$_2$
molecules. The flux has been previously calibrated using TPD by
determining the O$_2$ exposure time required to saturate the O$_2$
monolayer on np-ASW ice \citep{kimmel-2001-1}.

\item
The p-ASW substrate kept at 10~K is then exposed to varying amounts
of cold D atoms (50~K) ranging from 0~exposed monolayer (EML) to
2~EML. We use EML and not directly ML because in the case of light
molecules the sticking coefficient is not necessarily unity. Thus
less molecules are adsorbed on the surface than what the surface has been exposed to. 
Moreover in the case of p-ASW ice more
than 1~ML of D$_2$ can be adsorbed on the surface, and still the
coverage can be low, since the effective surface area is very large
due to the 3D conformation of the substrate \citep{Amiaud2006}.

\item
Finally, the QMS is placed in front of the surface and TPD mass
spectra of D$_2$ and O$_2$ are recorded simultaneously. All TPDs are done with a
linear heating rate of 10~K/min starting from 10~K up to a
temperature above which the O$_2$ molecules have completely desorbed
($\sim$65~K).
\item
The same series of experiments are also done by depositing D and D$_2$ followed by O$_2$.
\end{itemize}

%----------------------  Figure  -----------------------------
\begin{figure}
\resizebox{\hsize}{!}{\includegraphics{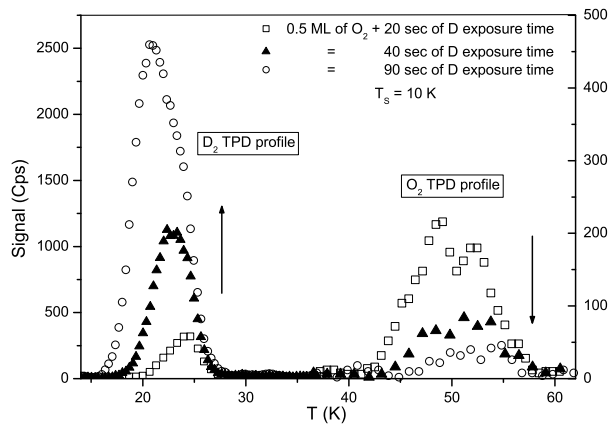}}
\caption{TPD profiles of D$_2$ and O$_2$ after deposition of 0.5 ML of O$_2$ and followed by different deposition times of D (20 seconds (squares), 40 seconds (triangles) and 90 seconds (circles))} \label{tpd}
\end{figure}
%----------------------  End figure  -------------------------

%----------------------  Figure  -----------------------------
\begin{figure}
\resizebox{\hsize}{!}{\includegraphics{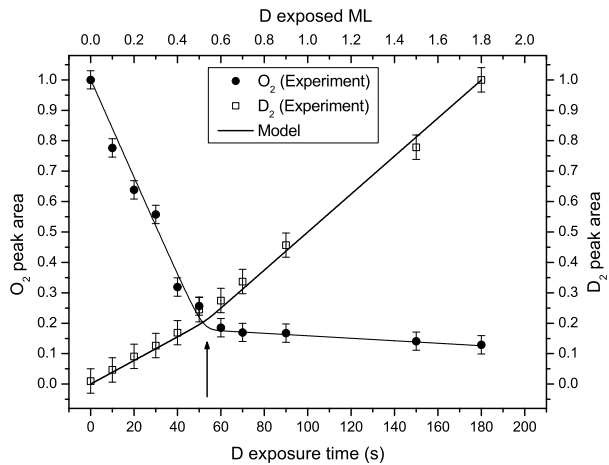}}
\caption{Normalised O$_2$ (plain circles) and D$_2$ (empty squares) TPD peak areas following different D
exposures. The O$_2$ normalisation was made with respect to the TPD peak area corresponding to 0.5~ML of O$_2$ alone and the D$_2$ normalisation was made with respect to the TPD peak area corresponding to 1.8 EML of D. O$_2$ and D both deposited at a surface temperature of 10~K. The  O$_2$ and D both deposited at a surface temperature of 10~K. The solid line represents the model results (see text).} \label{o2-d2}
\end{figure}
%----------------------  End figure  -------------------------

%%%%%%%%%%%%%%%%%%%%%%%%%%%%%%%%%%%%%%%%%%%%%%%%%%%%%%%%%%%%%%%%%%%%%%%%%%%%

\section{Results}\label{sectres}

\subsection{20 ML p-ASW substrate}\label{sect20ml}
As a first result we find that O$_2$ is only sensitive to D and not to D$_2$.
When D is deposited on the surface and then exposed, immediately after that,  to O$_2$, we find that the amount of desorbed O$_2$ is constant (within the experimental error bar (<10\%)) irrespective of the D amounts.
This series of experiments was done in order to rule out the possibility that D atoms were mobile only during the heating phase of the TPD. Indeed these results demonstrate that D atoms that adsorb on the ice surface diffuse immediately and either react with each other or desorb before O$_2$ is deposited. The amounts of desorbing O$_2$ are the same regardless of the amounts of D atoms deposited on the surface. Hence, in what follows, we restrict ourselves to experiments where O$_2$ is deposited first and then followed by various D exposure times.

Fig.\ref{o2-d2} displays the normalised TPD peak areas of
O$_2$ and D$_2$, as a function of D exposure time
(lower axis) or D EML (upper axis). Fig.\ref{tpd} and Fig.\ref{o2-d2} show clearly that the quantity of desorbing O$_2$
decreases with the increase of the exposed dose of D. A sudden change in the
slope of the monotonic decrease occurs after $\sim0.5$~EML of D exposure. At first, the drop of O$_2$
that desorb from the surface occurs rapidly, then it
becomes much slower. On the other hand, as expected, the amount of
D$_2$ desorbing from the p-ASW surface increases with
the exposure time of D (Fig.\ref{o2-d2}). Yet, in this case too,
we observe that a change in the slope of the D$_2$ desorption raise
takes place almost exactly after $\sim0.5$~EML of D exposure. Here the
increase of D$_2$ TPD peak areas is slightly accelerated after the slope change .

These observations can be explained as follows: as D atoms are deposited onto the p-ASW sample, two competing mechanisms occur, namely O$_2$ destruction and D$_2$ formation. If D
atoms are mobile, they are likely to encounter either another D atom
to form D$_2$ (D+D formation) or an O$_2$ molecule and react 
with it (O$_2$ destruction).
It is significant that in both cases of O$_2$ and D$_2$ desorptions, 
the change in slope occurs after the same amount of D
exposure time (see arrow in Fig.\ref{o2-d2}). This suggests that after $\sim50$~s (0.5 EML) of D
irradiation, when most O$_2$ has
been destroyed, a greater number of D atoms are available to form
D$_2$. The presence of a rapid regime of O$_2$ destruction (80\% of initial
dose destroyed by 0.5 EML of D) and of a slow regime (for the remaining 20\% of O$_2$ initial dose) can be ascribed
to the non perfect overlapping of the two beams. It has been checked that the two beams have in fact $\sim 80\%$ overlap on the surface.

\subsection{250 ML p-ASW substrate}\label{sect250ml}

\begin{figure}
\resizebox{\hsize}{!}{\includegraphics{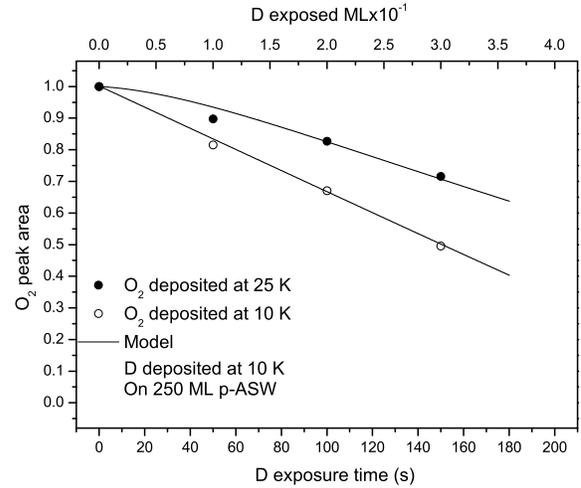}}
\caption{Comparison between normalised TPD peak areas of 0.5~ML of
O$_2$ deposited at 10 K (blank dots) and 25 K (black dots) and then
irradiated with 0, 50, 100 and 150 seconds of D atoms. The solid line represents the model results (see text).} \label{250ml}
\end{figure}

We then did the same experiment using a thicker p-ASW ice film in order to check whether the
destruction of O$_2$ on a 20~ML p-ASW ice substrate is
actually due to mobile D atoms being able to scan the porous surface. This
time, on 250~ML p-ASW films, we deposited O$_2$
at a higher temperature (25~K), to favour O$_2$ mobility and
to have O$_2$ not only adsorbed on the surface of the film
but also deeper into the porous structure.
The results are shown in Fig.\ref{250ml}. Plain circles 
represent the yield of desorbed O$_2$ after cooling the surface down to 10~K
and then irradiated with 0, 50, 100 and 150 seconds of D exposure at 10 K, while 
empty circles show the case of desorption of O$_2$ initially deposited at 10~K 
and then exposed to the same D exposure times.
First, we notice in these experiments that the O$_2$ destruction rate
is much slower than in the case of a 20~ML p-ASW substrate. We then only 
observe the first part of the curve in Fig.2 at a slower rate. This
can be due to the fact that the roughness is greater in the case of a
250~ML substrate. The structures (holes, pores, piles...) are more
pronounced. The total accessible surface is also higher and subsequently the average density of
O$_2$ is reduced. In this case D$_2$ formation is more favoured. 
Secondly, we find that if O$_2$ is deposited at 25 K, the
destruction rate is even lower. This can be due to the fact that at 25 K
O$_2$ is more mobile on the surface of the ASW ice than at 10~K. Therefore O$_2$ molecules are spread over all the accessible porous surface. O$_2$ density is again lower and D$_2$ formation is favoured. We also notice the disappearance of the elbow that was present in Fig.2 at 0.5 EML of D. It is due to the fact that we should have
exposed the surface to longer times of D atoms in order to reach
80\% of destruction of the adsorbed O$_2$ to observe it. Finally we have checked that all the O$_2$ deposited at 25 K disappears, after very large exposure times (600 seconds), even
on the 250 ML p-ASW film.
%----------------------  Figure  -----------------------------
%----------------------  End figure  -------------------------
%%%%%%%%%%%%%%%%%%%%%%%%%%%%%%%%%%%%%%%%%%%%%%%%%%%%%%%%%%%%%%%%%%%%%%%

\section{Model and data fit}\label{model}

A very simple model based upon the formalism of \cite{katz} was used to fit the experimental results already obtained and shown in Fig. 2. If F$_{tot}$ is the initial beam flux of D$_2$, after the dissociation F$_{tot}$ is divided into a flux of D, F$_D$, and a flux of D$_2$, F$_{D2}$. Both fluences on the surface are determined knowing the sticking probabilities of D$_2$ and of D, S$_{D2}$ and S$_{D}$ respectively, and the dissociation rate $\tau$ at a beam temperature of 50 K. The fluxes are given by:
\begin{center}
F$_D$ = 2$\tau$P$_D$F$_{tot}$
and
F$_{D2}$ = (1 - $\tau$)P$_{D2}$F$_{tot}$
\end{center}
Taking into account the porosity of the ice surface, the surface density of D, D$_2$ and O$_2$, and depending on the diffusion barrier E$_{diff}$ of D atoms on the surface, this model simulates the number of D + D reactions as well as the number of D + O$_2$ reactions.
As Fig. 2 and 3 clearly show, our simple model is able to reproduce the experimental results under the same conditions, namely, dissociation efficiency of the D$_2$ beam, sticking probabilities of D and D$_2$ (measured at a given beam temperature, Matar et al. \textit{in preparation}), and porosity of the ice. A best fit of the experimental data has given a diffusion barrier of 22$\pm$2 meV.

\section{Discussion}\label{sectdisc}
In the previous section we have proposed a direct interpretation of our 
experimental data assuming that D is mobile at 10 K. But other 
hypothesis should be also discussed in the light of these new 
experimental data, in particular the possible thermally induced mobility of 
the atoms. This process implies that atoms are not mobile at 10 K 
and that they require higher temperature to move and react.
If atoms are not mobile at 10 K, they should stay in the vicinity of 
the external surface, especially under our experimental conditions 
where D atoms have a kinetic temperature of 50 K (in the beam) 
which reduces considerably the mobility of D atoms resulting from 
the hot atom (Harris-Kasemo) mechanism.

The total number of D atoms sent to the surface can be quite high 
(2 10$^{15}$cm$^{-2}$, i.e. 2 ML) which corresponds to a value of full coverage 
of the external surface. Consequently, if atoms are not mobile, the 
surface density becomes high, and two mechanisms invoked in the literature 
should become non negligible. One is the direct Eley-Ridel formation 
mechanism for D$_2$ (an adsorbed atom forms a molecule with an atom 
incoming from the gas phase, a fraction of the formed molecules is then 
directly released in the gas phase), the second is the rejection 
mechanism \citep{perets2005} (adsorbed atoms prohibit atoms from the gas phase to stick 
on the surface). In both cases a fraction of D or D$_2$ evaporates during the exposure and therefore does not desorb later on during the heating ramp. If so, D$_2$ should not be proportional to the exposure, which is not the case in Fig.2 where we observe proportionality.
We can thus conclude that either the surface density does not increase 
to reach a value of the coverage close to unity, either the two cited mechanisms 
(Eley-Ridel and rejection mechanism) are not efficient. It is really 
unlikely that the Eley-Ridel mechanism does not occur, therefore 
density would not increase, implying that D atoms enter the pores.

Calculations \citep{buchczerminski} show that the adsorption 
energy of H on a water cluster and the height of the energy barrier 
between two adjacent sites are both described by large distributions. 
This means that the mobility is site dependent. Some deep adsorption 
sites can bind atoms more than others and release them less easily in the gas phase. 
Thus it is possible that a fraction of the atoms are not mobile at 10 K, on the 
time scale of the experiment, and require thermal activation to react. 
Our experiments just show that globally the D atoms are able to penetrate 
in the porous structure of the ice. In fact, we find a non-infinite D atom mobility E$_{diff}$ = 22$\pm$2 meV, and even if the model is based upon pure thermal hopping, this diffusion barrier physically means that the mean residence time before hopping is $\sim$12 ms. Our results do not show if this mobility is dominated by thermal hopping or tunnelling. More experiments should be done at lower surface temperature (below 8 K) to fully understand how this mobility occurs. 
Another aspect of these experiments concerns the efficiency of the 
O$_2$ + D reaction in comparison with the D + D reaction. We can see that 
O$_2$ + D $\rightarrow$ O$_2$D is a very efficient reaction \citep{watanabe2008, Ioppolo2008}, because the proportion of D required 
to destroy O$_2$ is close to 1, at the accuracy of our fluxes estimation 
($\sim$50\%), because O$_2$+D is believed to have a low energy barrier \citep{walch1988}. 
In our experiments the 
fact that the O$_2$+D reaction seems to be favoured in comparison with the D+D reaction when 
the exposition is low does not signify that the D+D reaction is less 
efficient because of an activation barrier. Indeed, our model assumes that both D+D and O$_2$+D reactions have 100\% efficiency. Once again, if atoms are 
mobile, the density of D atoms stays low because they are consumed by 
the O$_2$ population. It is only when the density of D atoms is equal to 
that of O$_2$ atoms that the production of D$_2$ can be effective. This 
consideration allows us to conclude that at any moment in our experiment, 
the density of D atoms is low, initially because they are 
consumed by O$_2$ and subsequently because they form D$_2$.

At the end of the experiment, when the ice is sublimated we 
detect a high proportion of HDO and not D$_2$O because of the isotopic exchange \citep{Smith1997}. Therefore O$_2$ is one of the possible precursor for water formation. Finally these experiments show that no O$_2$ should survive on grains in 
dark interstellar clouds. Apart from specific detection limitations, it 
can be a fundamental reason of the non detection of O$_2$ on icy grains in 
dark clouds. 

\section{Conclusions}\label{sectconc}

We have reported the results of experiments on the mobility of
deuterium at 10 K on porous ASW ice surfaces under
interstellar conditions, using a temperature programmed desorption
technique. Beams of O$_2$ and D were
irradiated on the surface of an p-ASW ice film and the mobility of D atoms at 10 K was investigated via their property of reacting with the O$_2$ molecules. By using a simple rate equation model we found that the D atoms diffusion barrier at 10 K is 22$\pm$2 meV. Our experimental results demonstrate that D atoms are mobile on ASW ice
surfaces at 10 K, and validate the chemical models that hypothesized
the mobility of H atoms at 10 K.

%----------------------  Acknowledgements  -------------------

\begin{acknowledgements}
The authors would like to thank L. E. Kristensen, V. Pirronello and O. Biham 
for fruitful discussions. We acknowledge the support of the
national PCMI program founded by the CNRS, as well as the strong
financial support from the Conseil Régional d'Ile de France
through SESAME programs (E1315 and I-07-597R), the Conseil Général du Val d'Oise and the Agence Nationale de la Recherche.
\end{acknowledgements}

%----------------------  Bibliography  -----------------------

\bibliographystyle{aa}
\bibliography{bibliography1}

\end{document}